# Online Social Networks and Terrorism 2.0 in Developing Countries

Fredrick Romanus Ishengoma

*College of Informatics and Virtual Education, The University of Dodoma, Dodoma, Tanzania.*

*ishengomaf@gmail.com*

**Abstract**

The advancement in technology has brought a new era in terrorism where Online Social Networks (OSNs) have become a major platform of communication with wide range of usage from message channeling to propaganda and recruitment of new followers in terrorist groups. Meanwhile, during the terrorist attacks people use OSNs for information exchange, mobilizing and uniting and raising money for the victims. This paper critically analyses the specific usage of OSNs in the times of terrorisms attacks in developing countries. We crawled and used Twitter's data during Westgate shopping mall terrorist attack in Nairobi, Kenya. We then analyzed the number of tweets, geo-location of tweets, demographics of the users and whether users in developing countries tend to tweet, retweet or reply during the event of a terrorist attack. We define new metrics (reach and impression of the tweet) and present the models for calculating them. The study findings show that, users from developing countries tend to tweet more at the first and critical times of the terrorist occurrence. Moreover, large number of tweets originated from the attacked country (Kenya) with 73% from men and 23% from women where original posts had a most number of tweets followed by replies and retweets.

## I.   Introduction

Online Social Networks (OSNs) example Twitter, Facebook, and YouTube have emerged as major means of communication for people in sharing and exchanging information on a wide variety of real-world events. As technology and Internet advances in developing countries, users from developing regions are also increasing in using OSNs (Reda, 2012). Numerous studies have shown that users from developing countries are spending a significant amount of time in OSNs (Karel 2007; Reda 2011). Another OSNs study (Pew Research Center, 2010) showed many people in developing countries do not go online, however when they have the opportunity to go online, they tend to use online social networking sites. For example, one-in-five Kenyans (19%) participate in online social networking, while just 5% use the Internet but do not participate.

Of the users that go online in developing countries, the demographics have shown a double-digit difference in gender gap where more men tend to go online compared to women (Pew Research Center, 2010; Intel, 2013). A recently study by Intel showed that on average, 23% fewer women than men are online in developing countries (Intel, 2013).

While the OSNs users can use the technology positively for communication and information exchange, yet the same technology can also be used negatively as a tool for terrorism 2.0. Terrorism 2.0 is the use of web 2.0 applications and semantic technologies in assisting or performing terrorist acts. With web 2.0, terrorist are able to connect them using social networks for sharing information, disseminate propaganda, recruiting new members and even planning an attack. For example, Al-Shabaab, a terrorist group based in Somalia, used Twitter during a Westgate terrorist attack in Nairobi to disseminate information and claiming responsibility of the attack.

Meanwhile, during terrorist attacks people have been using these OSNs to break the news, provide information with rich media content and uniting people for providing help and fundraising for the victims. For example, during the Westgate attack Twitter was used to mobilize the country for blood donation, money donation as well as keeping the peace. People used the hashtag #WeAreOne for uniting the country through updates and insights and raising money for the victims.

The aim of this paper is to analyze how OSNs is being used during the event of a terrorist attacks in developing countries along. We perform activity analysis of Twitter after the Westgate attack in Nairobi. Moreover, we present the OSNs challenges and opportunities in relation to terrorism 2.0. The rest of the paper is organized as follows: We start by providing related work in section 2. Section 3 is dedicated to the background of terrorism 2.0 and OSNs. Section 4 discusses Twitter in the era of Terrorism 2.0 with subsections on how people use Twitter and Twitter as a terrorist assistance tool. Data collection is discussed in Section 5. Section 6 discuses and analyses the findings of the study. We conclude in Section 7.

## II.    Related Work

In this section, we provide an overview of various studies that are related to our research work. A number of studies (Murthy, 2010; Mendoza, 2010, Stollberg, 2012; Nagar, 2012; Hossmann, 2012) have focused their work on the usage of OSNs during disasters like floods and earthquakes and in facilitating event detection and responses to emergency situations (Becker, Naaman, and Gravano 2011; Li, Sun, and Datta 2012; Li et al. 2012; Weng et al. 2011).

Weimann (Weimann, 2011) describes how the advances of technology has changed terrorist online communication from one-directional to bi-directional with interactive capabilities like chat-rooms, online social networking sites, video-sharing sites. (Veerasamy, 2012) represented the functions and methods that terrorists have come to rely on through the Information and Communication Technology (ICT) infrastructure. The discussion sheds light on the technical and practical role that ICT infrastructure plays in the assistance of terrorism.

(Gupta, 2011) attempt to characterize and extract patterns of activity of users on Twitter during a crisis based on the Mumbai terrorist attack. The study attempted to characterize and extract patterns of activity of users on Twitter during a terrorist situation. Another close related work is by Oh and Agrawal, (Oh, 2011). They analyzed the Twitter stream during the 2008 Mumbai terrorist attacks. The study applied social awareness theory to show how information available on OSNs during the attacks aided the terrorists' decision making.

While some of the studies have the similar objectives with our study, our study remains unique as it focuses on a OSNs usage during terrorists attack in a developing country context.

### III.    Terrorism 2.0 and Online Social Networks

In this section we will provide an introductory overview of Terrorism 2.0 and Online Social Networks (OSNs) and how they are related.

*A. Terrorism 2.0*

As Information and Communication Technologies (ICTs) advances and transform how we live, it also changes the terrorism era. Advances in IT has brought a new era of Terrorism 2.0 where terrorist now make use of high advancement of IT to accomplish their goals. OSNs have now been used by terrorists for communication, recruiting new members, transmitting training videos and materials and propaganda.

For example, one terrorist website provide details on how to make Improvised Explosive Devices (IEDs), a bomb that can be made from locally available materials. Other terrorist websites have been distributing materials (tutorials and books) online on how to make poisons, bombs and conduct a terrorist attack. Some of the popular terrorist books that circulate online are The Mujahideen Poisons Handbook and The Anarchist's Cookbook.

*B. Online Social Networks*

Online Social Networking Sites (OSNs) are kind of social media that gives ability online user to make a profile with his information and enable users to share and exchange messages and multimedia content (Boyd & Ellison, 2008). Some examples of popular social networks include Facebook, Twitter, YouTube and Flickr. They depend on peer-to-peer (P2P) networks that are cooperative, distributed, and community driven. Within the OSNs users can create groups for people with common interest to share information and communicate privately.

Recently, terrorists have changed from its traditional ways of communicating to adapt the use of OSNs as their platform of communication. OSNs has opened a new way to terrorists enabling them to communicate easily and more secure, disseminate information, propaganda and training materials, organize attacks, recruiting new members and to seek sympathy to the public for their actions worldwide.

### IV.    Twitter in the Era of Terrorism 2.0

With registered users more than 500 million worldwide as of October 2013, Twitter is one of the most popular online social networks available. Twitter offers exchange of short messages called Tweets of up to 140 characters long between its users.  Hashtags are words prefixed with '#' used by Twitter users to describe a subject (e.g. #BreakingNews). Common practice of responding to a tweet has evolved into well-defined mark-up culture: RT stands for Retweet. A retweet is a message from one user that is forwarded by a second user. The retweet mechanism empowers users to spread information of their choice beyond the reach of the original tweet's followers. A message from one Twitter user that is a response to Twitter user's message is called a Reply. Being a follower on Twitter means that the user receives all the tweets from those the user follows. We opted to use Twitter as our choice of OSN for this study due to its availability of data.

### A. How people use Twitter?

Studies have shown that people use Twitter for different purposes (Java, 2007; Ramage, 2010; Naaman, 2010) studied the aims of 94,000 Twitter users with more than 1.3 million tweets and concluded that users make use of Twitter in the following major areas:

(i)      Chatting. Most Twitter users are for discussion of day-to-day activities.

(ii)     Conversations. Some of the Twitter users engage in the conversation using replies mechanism.

(iii)    Information exchange/Sharing. Twitter users share and exchange information through Twitter posts contains URLs.

(iv)     News. Reporting and commenting on breaking news and latest news is another type of people's usage in Twitter.

### B. Twitter as a Terrorist Assistance Tool

OSN is a suitable platform for terrorists because of its easy access, little control from the governments, worldwide audience, anonymity (via fake profiles), fast flow of information, cheap development of OSN websites and multimedia platform (the capability to combine text, graphics, audio, video and downloading). With the current highly development of Information and Communication Technologies (ICTs) terrorist uses OSN for the following purposes:

(i) *Information exchange.* Terrorist use OSN to communicate with worldwide audience than it was a decade before. Terrorists have become sophisticated in using anonymous/secret communications using encryption, steganography and anonymity softwares. Example, Al-Qaeda members have been using encryption softwares "Mujahedeen Secrets 1" and  "Mujahedeen Secrets 2" to encrypt and secure their email communications.

(ii) *Recruiting and Training.* A report by the institute of homeland security of the George Washington University (Homeland Security Institute, 2009) relays that the Internet has become crucial instrument in the hands of terrorists to spread their messages and recruit new supporters. Nowadays terrorists' recruitment and training can be successfully offered via OSNs easily and anonymously unlike before, which required the physical meetings of trainees and trainers. Recruits are passing through a series of tests in password protected websites and restricted chat rooms before accepted and joining the terrorist group (Gerwehr, 2006). Tactics used includes the integration of terrorist acts in cartoons and music videos to attract the minors into terrorism. Also, video games that involve the acts of terrorism like mass suicide attack.

(iii) *Planning attacks.* OSNs are used by terrorists to plan an attack due to its secure communication and fast message channeling.

(iv) *Fundraising*. Through wire transfer and email address the groups are able to conduct a fundraising in OSNs.

(v) *Cyber-attack*. Cyber-attack refers to the use of computer networking tools to attack other computer networks or national communications systems like government operations, transportation, and energy.

(vi) *Propaganda*. Though these OSN have played a great role in the society, at the same time the terrorists is using them for their propaganda dissemination. These propaganda deals with delivering ideological, clarifications, explanations or campaign of terrorist actions. These can involve messages, demonstrations, magazines, audio and video footage of violent acts.

On 16 March 2013, the propaganda arm of al-Qaeda, the Andalus Foundation, created a twitter account (@Andalus_Media). The account has gained more than 14,500 followers while following 7 people inclusing the Somali terrorist group Al Shaabab. Al-Shabaab (Kimunguyi (2010), Ibrahim (2010)) also known as Harakat al-Shabaab al-Mujahideen (HSM) started their Twitter account in English as HSM Office on December 2011 using the Twitter handle '@HSMPress' (https://twitter.com/#!/HSMPress). HSMPress is the press branch of Al-Shabaab (HSM). Since then they have been using Twitter to exaggerate their military accomplishments.

Al-Shabaab has been in battle with Kenya (due to presence of Kenyan Army in Somalia) that has led to Westgate attack in September 2013 Nairobi. The terrorist group's Twitter account has been repetitive suspended by Twitter due to violation of terms of service. Twitter's terms of service state that users "may not publish or post direct, specific threats of violence against others." However each time their account is suspended, they create new accounts with some slight variations on the same name such as @HSMPRESS1, @HSM_PressOffice, @HSM_PROffice and @HSM_PR.

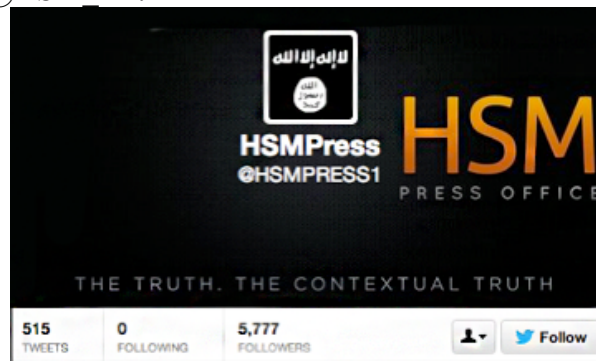

**Figure. 1.** Cover of Al-Shabaab's English Twitter page as of 13 August 2013.

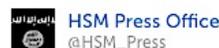

HSM Press Office
@HSM_Press

The attacks are just retribution for the lives of innocent Muslims shelled by Kenyan jets in Lower Jubba and in refugee camps
#Westgate

**Figure. 2.** Example of Al-Shabaab's Tweet during the Westgate attack.

### V.    Data

Our case study is the terrorist attack of Westgate shopping mall in Nairobi, Kenya on 21st September 2013. The attack lasted until 24th September resulting in at least 72 deaths and over 200 wounded people (Raidió Teilifís Éireann, 2013). The Islamist group Al-Shabaab claimed accountability for the terrorist attack.

We used Twitter search API during the Westgate attack period when the text string '#Westgate' was most active and Topsy (Topys, 2013), a certified Twitter partner that maintains the world's largest index of tweets, numbering in hundreds of billions, dating back to May 2008. We crawled the data of the popular Twiter #hashtags that was used during the Westgate attack namely; #Westgate. The #Westgate hashtag was used in tweeting by both the terrorists and the civilians. We analyzed data between 21st and 27th to obtain the clear picture of the patterns. These data are analyzed to explore the following metrics:

     (i)      Number of tweets during the terrorist attack.

     (ii)      Whether users tend to tweet, retweet or reply in Twitter during the event of Westgate terrorist attack.

     (iii)      What countries had the highest frequency of tweets during the Westgate terrorist attack.

     (iv)      What are the demographics of the Twitter users during the Westgate terrorist attack and

     (v)      What were the reachness and impressions of the tweets during the Westgate terrorist attack?

### VI.    Findings and Discussion

In this section we present our findings of the study and discussion. We begin by examining the number of tweets that were sent during the #Westgate attack. While there were many other hashtags used to discuss the terrorist attack in Twitter posts (like #Nairobi, #Kenya, #Terrorist), we chose the #Westgate hashtag since it was the most relevant one and standard one. We considered the number of tweets that uses only the hashtag 'Westgate'.

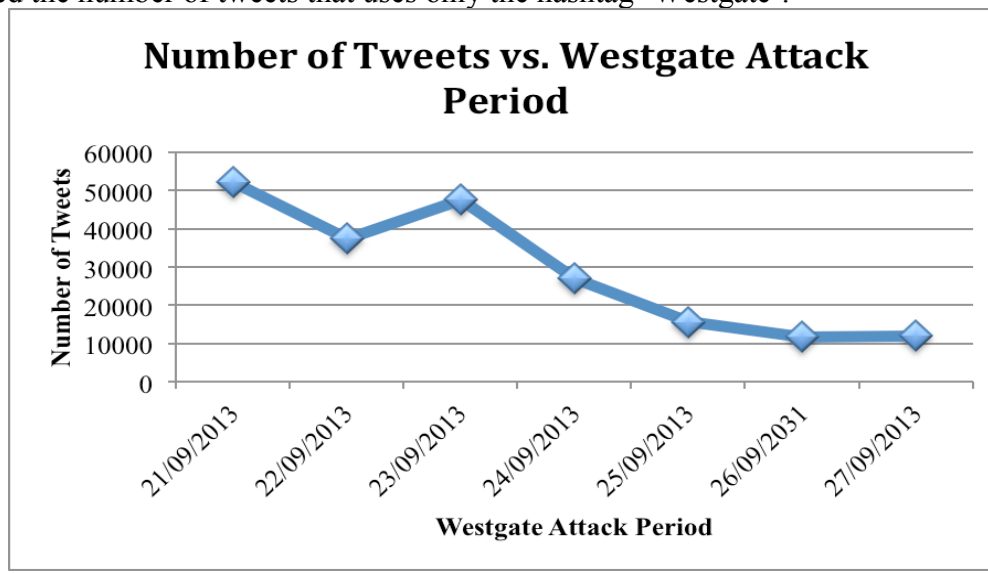

**Figure. 3.** Number of tweets during the #Westgate attack.

Figure 3 shows the number of tweets tweeted using #Westgate hashtag during the Westgate attack period between 21st September.  The number of tweets is observed to be high during the first hours of the terrorist attack and when the terrorist attack is in its peak. This is due to the Twitter's real time nature i.e. the behaviour of Twitter users to post about events as they are happening. The tweets are observed to decrease as the terrorist attack ends.

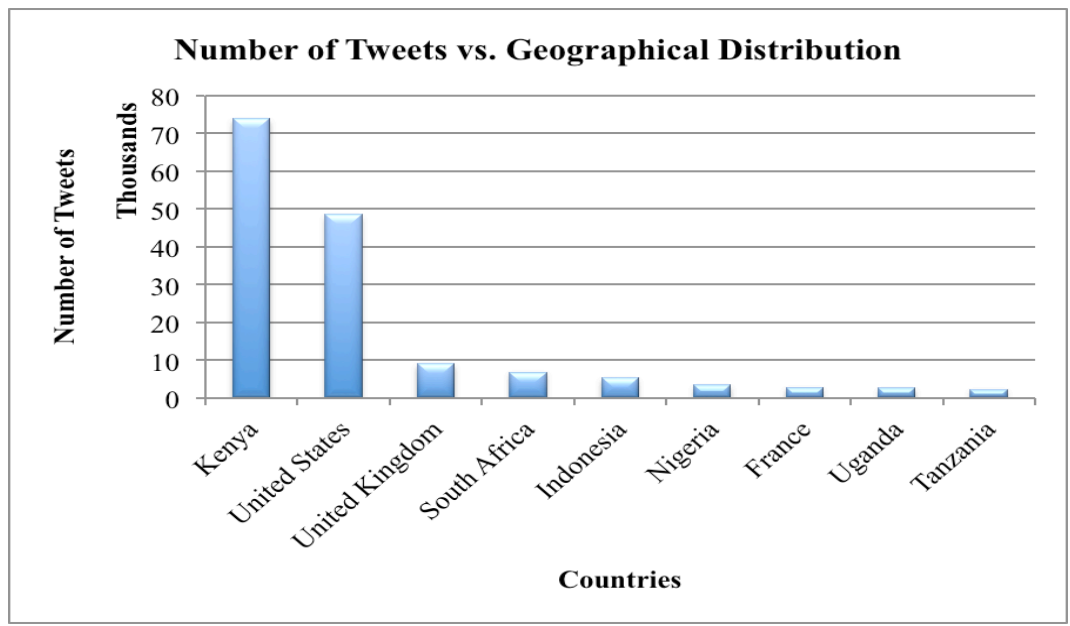

**Figure 4**. Geographical distribution of tweets during the #Westgate attack.

Figure 4 shows the geographical distribution of tweets during the Westgate attack. From the figure it is observed that most of the tweets came from the developing countries. Kenya being the countries tweeted mostly about the attack, this could be because it is the attacked country. Also, a significant number of tweets is shown to have originated from US and UK.

We then analyzed the Twitter data to find out, how many tweets were original posts, retweets or replies using the keyword #Westgate.

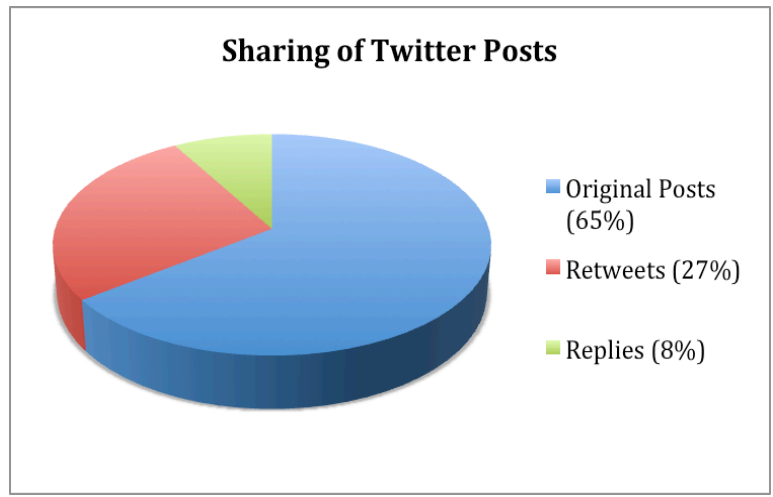

**Figure 5**. Sharing of tweets (original, retweets and replies) during the #Westgate attack.

Figure 5 shows the sharing of Twitter posts during the Westgate attack. From the figure it is observed that most of the tweets during the attack were original posts (65%), followed by retweets (27%) and replies (8%).

In figure 6 we analyzed the reach and impression of #Westgate tweets during the terrorist attack. We define reach, as the number of unique followers that a user has that is, the unique people who a tweet could potentially get to. We define impression as the number of times a user posts that is, how many times followers would see these posts. Let N be the number of followers a Twitter user has and t the number of times a tweet is tweeted. Reach is given by N whereby Nt gives impression. For example, if a Twitter user has 300 followers In figure 6 we analysed the reach and impression of #Westgate tweets during the terrorist attack. We define reach, as the number of unique followers that a user has that is, the unique people who a tweet could potentially get to. We define impression as the number of times a user posts that is, how many times followers would see these posts. Let N be the number of followers a Twitter user has and t the number of times a tweet is tweeted. Reach is given by N whereby Nt gives impression. For example, if a Twitter user has 300 followers and tweet 3 times then tweet's reach is 300 and tweet's impression is 900 (since 300 followers saw the tweet thrice).

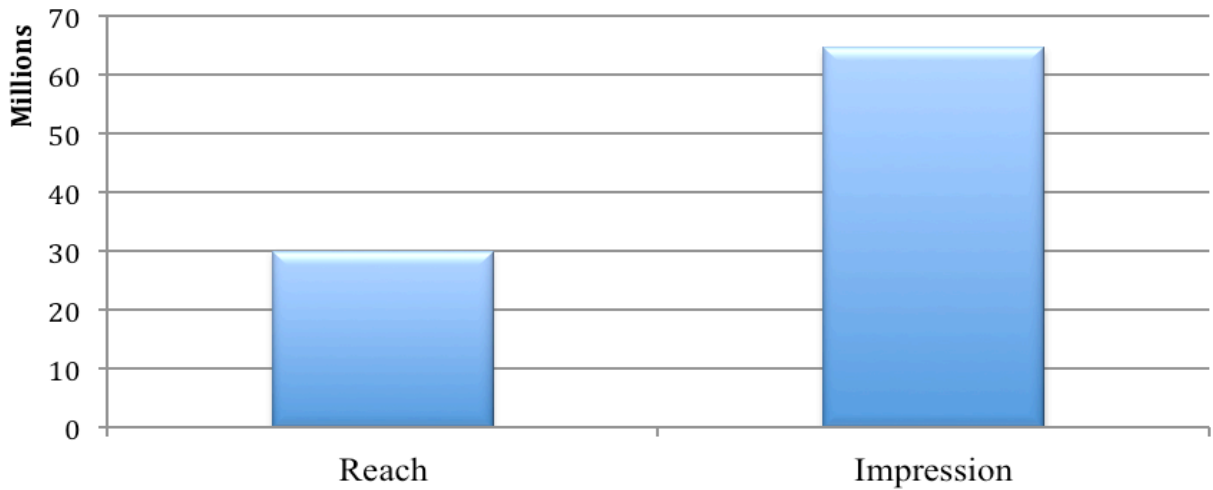

**Figure 6.** Reach and Impression of tweets during the #Westgate attack.

We then analyzed the demography of the Twitter posts during the #Westgate attack. In Figure 7 we could see the percentage of male and female who tweeted about the terrorist attack using #Westgate hashtag. Males are observed to tweet more than females. 73% of the #Westgate tweets are observed to be from males while 27% of the #Westgate tweets are from females. This could be due to the gender gap in developing countries where according to International Telecommunication Union (ITU, 2013), 16% fewer women than men use the Internet in developing countries.

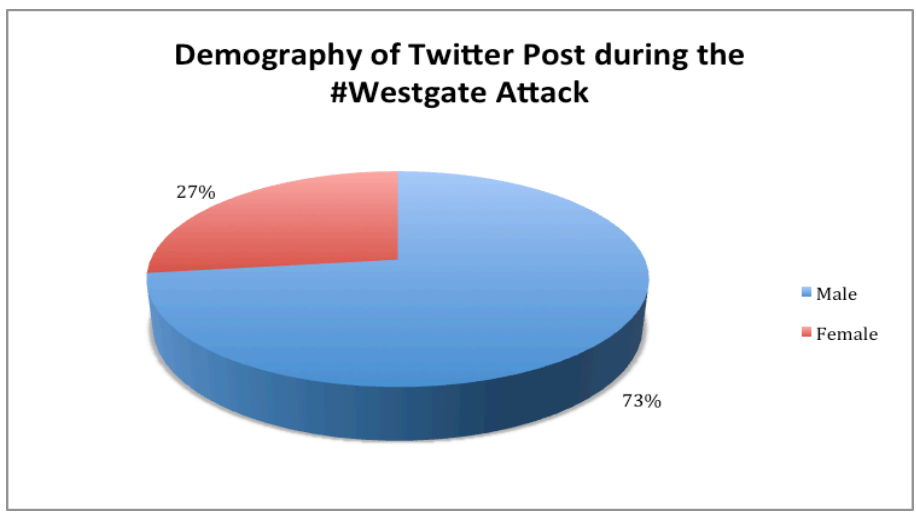

**Figure 7.** Demography of Twitter posts during the #Westgate attack.

VII.     **Conclusion**

In this paper, we analyzed the usage of Online Social Networks (OSNs) in the event of a terrorist attack with a case study of a Westgate shopping mall attack in Nairobi, Kenya. We used different metrics like number of tweets, whether users in developing countries tended to tweet, retweet or reply, demographics, geo-location and we defined new metrics (reach and impression of the tweet) and presented their models. While the developing countries are faced by many limitations in using OSNs such as unreliable power and poor Internet connection, still the study finding challenges the traditional media of reporting during disasters like terrorists attacks. We recommend centers globally to make full use of the OSNs for crisis communication in order to save more lives during such.